\documentclass[aps,epsfig,prb,twocolumn]{revtex4}
\usepackage{amsmath}
\usepackage{graphicx}

\def\nn{\nonumber}

\begin{document}

\title{Massive Dirac Fermions Signal in Raman Spectrum of Graphene}

\author{Ken-ichi Sasaki}
\email{sasaki.kenichi@lab.ntt.co.jp}
\affiliation{NTT Research Center for Theoretical Quantum Physics and NTT Basic Research Laboratories, NTT Corporation,
3-1 Morinosato Wakamiya, Atsugi, Kanagawa 243-0198, Japan}

\date{\today}

\begin{abstract}
 Massless Dirac fermions in graphene can acquire a mass through
 different kinds of sublattice-symmetry-breaking perturbations, and 
 there is a growing need to determine this mass using a conventional method.
 We describe how the mass caused by a staggered sublattice potential
 is determined using Raman spectroscopy and
 explain the mechanism in terms of the pseudospin polarization of massive Dirac fermions.
\end{abstract}

%\pacs{73.20.-r, 73.40.-c, 72.80.Vp}
\maketitle

The mass of Dirac fermions in graphene is a broad subject 
relating to various aspects of physics from 
intriguing phenomena 
such as the quantum Hall effect~\cite{haldane88} and quantum spin Hall effect~\cite{Kane2005}
to the device application of graphene
such as the band-gap engineering.~\cite{Novoselov2007} 
Dirac fermions in conventional graphene are massless.~\cite{Novoselov2005,zhang2005} 
However, they acquire different kinds of masses 
depending on the patterns of the symmetry breaking of 
the equivalence between two carbon atoms 
in the hexagonal unit cell (A and B atoms).
For example, 
an inversion-symmetry-breaking potential energy $+m$ on A atoms and $-m$
on B atoms~\cite{semenoff84,Hunt2013a,Kindermann2012a}
is an essential perturbation ($m\sigma_z$)
that opens the band-gap $E_g=|2m|$ and a theoretically estimated $E_g$ of
approximately 50 meV is reported for graphene placed on hexagonal boron nitride (h-BN).~\cite{Giovannetti2007,Fan2011}
Strikingly, the mass ($m$) can be a positive or negative number 
depending on the sign of the potential energy of one sublattice, and 
a spatial change in the sign of $m$ is responsible for topological phenomena.
It is considered that a lattice mismatch between graphene and h-BN may
induce a domain wall (where $m$ vanishes),
along which a topologically protected one-dimensional conducting channel
of Weyl fermions is formed as midgap states.~\cite{Semenoff2008}
%This topological aspect of the mass indicates that the mass of Dirac fermions 
%is not another way of saying the band-gap.
The sign of the mass is combined with valley ($\sigma_z \tau_z$) and
spin ($\sigma_z \tau_z s_z$) degrees of freedom, which are essential
ingredients of a topological insulator.~\cite{Kane2005}
There is a growing need to identify mass.
%Besides, recent experiment with angle-resolved photo-emission
%spectroscopy reports that ordered buffer graphene grown epitaxially on
%SiC shows the spectrum of Dirac fermion with a band gap of 0.5
%eV.~\cite{Nevius2015}

In this paper, we show the effect of the mass 
caused by an inversion-symmetry-breaking potential in graphene
on the self-energies of Raman active phonons, namely 
the $G$ and $2D$ ($G'$) bands.~\cite{Ferrari2007,Malard2009}
The phonon self-energy arises from the interaction between optical
phonons and Dirac fermions,
and signifies various aspects of the electronic state.
We will show that the $G$ band exhibits the characteristic signal of the
mass, while the $2D$ band generally does not.

When exploring the physics related to mass, 
we must always pay attention to the position of the Fermi energy
($E_F$) because the phonon self-energies depend on it.
Moreover, $E_F$ varies even unintentionally during the graphene
samples fabrication process in a manner that depends on the substrate
condition.
It is known that for the massless Dirac fermions
$E_F$ is determined by the properties of the self-energies of the $G$ and $2D$ bands;
when $E_F$ is close to the Dirac point ($E_F\simeq 0$),
the $G$ bandwidth broadens,~\cite{yan07,pisana07,das08nature,lazzeri06prl,ando06-ka}
by contrast, the $2D$ bandwidth sharpens.~\cite{das08nature,chen11}
Thus, our task is to obtain the phonon self-energies as a function of $E_F$ and $m$.
The two controllable variables make graphene a fascinating playground 
in which to simulate a novel aspect of quantum electrodynamics.
The ground state (or Dirac sea) of our universe cannot be changed from
charge neutrality.
In contrast, the ground state of graphene is controllable.
Confirming the charge neutrality of massive Dirac fermions using Raman
spectroscopy is also a critical issue in relation to the observation of
the Weyl fermions that propagate at a domain wall.~\cite{Semenoff2008}

This paper is organized as follows.
In Sec.~\ref{sec:form},
we describe the model and formulation used to calculate the phonon self-energies.
In Sec.~\ref{sec:results}, 
we show the phonon self-energies for the $G$ and $2D$ bands, 
which are the main result of this paper. 
Our discussion and a summary are provided in Sec.~\ref{sec:discussion}.

\section{Model and Formulation}\label{sec:form}

The Hamiltonian of massive Dirac fermions is defined as
\begin{align}
 \hat{H} = 
 \begin{pmatrix}
  +m & \hbar v(\hat{k}_x - i\hat{k}_y) \cr
  \hbar v(\hat{k}_x + i\hat{k}_y) & -m
 \end{pmatrix},
 \label{eq:HDirac}
\end{align}
where $v$ and $\hat{k}_i=-i\partial/\partial x_i$
are the velocity and wavevector operator, respectively.
When $m$ is uniform in space, 
the energy eigenvalues of the conduction and valence bands are
$E_k \equiv \sqrt{(\hbar vk)^2 + m^2}$ and $-E_k$, respectively.
We assume a positive mass $m>0$ unless otherwise mentioned.
Using the variables $\theta_{\bf k}$ satisfying 
${\bf k}=k(\cos\theta_{\bf k},\sin\theta_{\bf k})$ and 
$\phi_k$ satisfying $\cos\phi_k=m/E_k$ and $\sin\phi_k=k/E_k$,
we express the wavefunction as the plane-wave multiplied by the pseudospin,
\begin{align}
 \psi_{v}({\bf k}) =
 \begin{pmatrix}
  e^{-i\theta_{\bf k}} \sin\frac{\phi_k}{2}  \cr -\cos\frac{\phi_k}{2}
 \end{pmatrix}, \ 
 \psi_{c}({\bf k}) = 
 \begin{pmatrix}
  e^{-i\theta_{\bf k}} \cos\frac{\phi_k}{2}
  \cr \sin\frac{\phi_k}{2}
 \end{pmatrix}.
\end{align}
The state at the bottom of the conduction
band has amplitude only on A-atoms because $\psi_c=(1,0)^{t}$ for ${\bf k}=0$.
Contrastingly, the state at the top of the valence band 
has amplitude only on B-atoms because $\psi_v=(0,1)^{t}$.
Thus, the low-energy states near the charge neutrality are highly
pseudospin polarized states and the low-energy interband transitions are
associated with the flip of the pseudospin, as shown in the inset of Fig.~\ref{fig:Gband}.

The self-energy of a phonon with wavevector ${\bf q}$ 
and frequency $\omega_{\bf q}$ is defined by
\begin{align}
 \Pi(m,E_F) = g_s g_v \sum_{s,s'} \sum_{\bf k}
 \frac{f_{E_F}(\varepsilon_{\bf k}^s)-f_{E_F}(\varepsilon_{\bf k+q}^{s'})}{\varepsilon_{\bf k}^s- \varepsilon_{\bf k+q}^{s'} +
 \hbar \omega_{\bf q} + i\epsilon}
 |M_{{\bf k+q},{\bf k}}^{s',s}|^2.
 \label{eq:selfenergy}
\end{align}
Here, $g_s$ ($g_v$) is spin (valley) degeneracy, and 
$(g_s,g_v)=(2,2)$ for the $G$ band and $(2,1)$ for the $2D$ band.
The electronic band structure is rewritten as $\varepsilon_{\bf k}^s (\equiv sE_k)$ 
where $s=1$ ($-1$) denotes the conduction (valence) band, and $\epsilon$ is a positive infinitesimal.
The Fermi distribution function 
$f_{E_F}(\varepsilon_{\bf k}^s)=\lim_{T\to 0}
(1+e^{(\varepsilon_{\bf k}^s-E_F)/k_B T})^{-1}$ 
is defined at zero temperature, where we can assume $E_F \ge 0$ without losing generality 
because $\varepsilon^{s}_{\bf k}$ is antisymmetric with respect to $s$.
$\Pi(m,E_F)$ is a complex number, 
and the broadening and energy shift of the phonon are given by
$-{\rm Im} \Pi(m,E_F)$ and 
${\rm Re} \Pi(m,E_F)$, respectively.
The term, $|M_{{\bf k+q},{\bf k}}^{s',s}|^2$,
is the absolute square of the transition amplitude of the process in which the
electron is transferred from $(s,{\bf k})$ to $(s',{\bf k+q})$ by
absorbing the phonon with ${\bf q}$.
Because of the momentum selection rule of the first order Raman process, 
the $G$ band consists of degenerate 
longitudinal optical (LO) and transverse optical (TO) phonons at ${\bf
q} \to 0$. 
We adopt a gauge theory framework of lattice
deformation
to calculate $M_{{\bf k+q},{\bf k}}^{s',s}$.~\cite{Sasaki2012b} 
The electron-phonon interaction is given by replacing 
$\hat{\bf k}$ in Eq.~(\ref{eq:HDirac}) with $\hat{\bf k}+{\bf A}({\bf r})$, 
where ${\bf A}({\bf r})$ represents the phonon fields where
we can set ${\bf A}({\bf r})=(A_x,0)e^{i{\bf q}\cdot {\bf r}}$.~\cite{sasaki08ptps}
By introducing the polar angle $\varphi_{\bf q}$
between the vector ${\bf q}$ and the $k_x$ axis,
the LO mode corresponds to $\varphi_{\bf q}=\pi/2$,
while the TO mode corresponds to $\varphi_{\bf q}=0$
because the LO mode satisfies $\nabla \cdot {\bf A}^{\rm LO}({\bf r})= 0$ and 
the TO mode satisfies $\nabla \times {\bf A}^{\rm TO}({\bf r})=0$.
The corresponding electron-phonon matrix element is
$M_{{\bf k+q},{\bf k}}^{s',s}
 = \psi_{s'}({\bf k+q})^{\dagger}
 \left\{ \hbar vA_x \sigma_x \right\} 
 \psi_s({\bf k})$.
It is straightforward to show that 
\begin{widetext}
\begin{align}
 |M_{{\bf k+q},{\bf k}}^{s',s}|^2 = 
 \frac{g}{2} \left[
 1-ss' \left\{ \frac{m^2-(\hbar v k)^2 \cos (2\varphi_{\bf q}-2\varphi) -(\hbar v)^2 kq \cos (2\varphi_{\bf q}-\varphi)}{E_k E_{|{\bf k+q}|}} \right\}
\right],
 \label{eq:M2}
\end{align}
\end{widetext}
where $\varphi$ denotes the polar angle between ${\bf k}$ and ${\bf q}$,
and the factor $g$ ($\equiv (\hbar v A_x)^2$) denotes the electron-phonon coupling
strength.
We calculate Eq.~(\ref{eq:selfenergy}) in the continuum limit by setting
$\sum_{\bf k}\to V/(2\pi)^2\int d^2{\bf k}$ and treat 
$gV/(2\pi \hbar v)^2$ as a parameter.
We will show $\Pi(m,E_F)$, which is divided by $\hbar \omega_{\bf q}
gV/(2\pi\hbar v)^2$.

\section{Phonon self-energies}\label{sec:results}

The analytical calculation of Eq.~(\ref{eq:selfenergy}) is
straightforward but lengthy.
We show it in the Supplement and focus on the calculated results here.
The energy of the $G$ ($2D$) band phonon $\hbar \omega_G$ ($\hbar
\omega_{2D}/2 \equiv \hbar \omega_{D}$) is approximately 0.2 eV (0.16 eV), 
which serves as a characteristic energy scale that governs the behavior of the self-energy.

\subsection{The G Band}

\begin{figure*}[htbp]
 \begin{center}
  \includegraphics[scale=0.55]{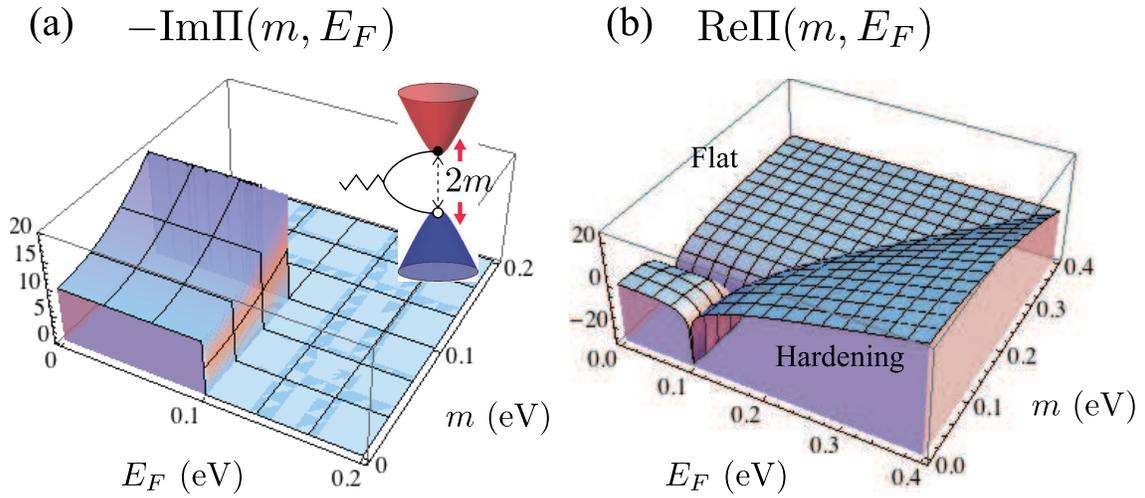}
 \end{center}
 \caption{(color online) 
 The spectrum broadening (a) and shift (b) of the $G$ band, 
 where $\hbar \omega_G=0.2$eV.
 The inset in (a) shows that the steep structure seen at $m=\hbar \omega_G/2$ 
 originates from pseudospin polarization caused by the mass.
 When $m=0$, $\Pi(0,E_F)$ reproduces the Kohn anomaly previously discussed for
 massless Dirac fermions.
 }
 \label{fig:Gband}
\end{figure*}

Figure~\ref{fig:Gband}(a) shows a three-dimensional plot of the calculated
broadening of the $G$ band,
\begin{align}
 -{\rm Im} \Pi(m,E_F)
 = & \pi^2 \left\{ 1+\left(\frac{2m}{\hbar\omega_G}\right)^2
  \right\} \theta_{\frac{\hbar\omega_G}{2}-m}
 \theta_{\frac{\hbar\omega_G}{2}-E_F}, 
 \label{eq:-imP}
\end{align}
where $\theta_x$ denotes a step function 
satisfying $\theta_{x\ge 0}=1$ and $\theta_{x<0}=0$.
The variables $m$ and $E_F$ are given in units of eV.
We note that in Eq.~(\ref{eq:-imP})
the $E_F$ dependence appears only through the step function 
$\theta_{\frac{\hbar\omega_G}{2}-E_F}$.
When $E_F \le \hbar\omega_G/2$ ($=0.1$eV), 
$-{\rm Im}\Pi(m,E_F)$ continues increasing as $m$ increases up to
$\hbar\omega_G/2$ at which the value takes its maximum which is double that at $m=0$.
A further increase in $m$ causes an abrupt change in the spectrum width,
namely, 
$-{\rm Im}\Pi(m,E_F \le \hbar\omega_G/2)$ vanishes when $m > \hbar \omega_G/2$.
This is because the broadening of the $G$ band is caused by the resonant decay
of the phonon into a vertical electron-hole pair 
(see the inset of Fig.~\ref{fig:Gband}(a)) and 
energy conservation forbids the phonon from decaying into an electron-hole
pair when $E_g > \hbar \omega_G$.
%In other words, the $G$ peak sharpens when $m \ge \hbar \omega_G/2$ and
%the spectrum width (observed when $m \ge \hbar \omega_G/2$)
%originates from the other effects besides the electron-phonon
%interaction, such as anharmonic phonon-phonon scattering.

The steep structure at $m=\hbar \omega_G/2$ in $-{\rm Im}\Pi(m,E_F)$
is unrelated to the density of states (DOS) but is related to 
the effect of pseudospin polarization induced by the mass. 
The DOS is proportional to $|E|\theta_{|E|-m}$ and has no van Hove singularity.
As shown in Sec.~\ref{sec:form}, 
massive Dirac fermions at the top of the
valence band and those at the bottom of the conduction band are
polarized in terms of pseudospin.
Because the electron-phonon interaction is proportional to Pauli's spin
matrix $\sigma_x$,~\cite{Sasaki2012b} 
the probability of decay into the interband electron-hole pair is maximum.
Indeed, the transition probability is known from Eq.~(\ref{eq:M2}) as
\begin{align}
 \lim_{\bf q\to 0}
 \frac{|M^{-s,s}_{\bf k+q,k}|^2}{g} = 1-\frac{(\hbar v k)^2}{(\hbar v k)^2+m^2}
 \begin{cases}
  \sin^2 \varphi \ \ ({\rm LO}) \cr
  \cos^2 \varphi \ \ ({\rm TO}).
 \end{cases}
\end{align}
Thus, when $m\ne 0$ the probability is maximum (unity) in the $k\to 0$
limit, while when $m=0$ the transition probability is a half of unity
determined from the average over $\varphi$, irrespective of the $k$ value.
The pseudospin of a massless fermion stays in the plane spanned by $\sigma_x$ and
$\sigma_y$, and is not polarized with respect to $\sigma_z$.
This is responsible for the well-known fact that for $m=0$
the broadening is invariant when $E_F < \hbar \omega_G/2$.~\cite{ando06-ka,lazzeri06prl} 
Note that when $E_F > \hbar \omega_G/2$,
the Pauli exclusion principle forbids the phonon from decaying into an
electron-hole pair.

%%%%%%%%%%%%%%%%%%%%%%%%%%%%
%%%%%%%%%%%%%%%%%%%%%%%%%%%%

Figure~\ref{fig:Gband}(b) shows ${\rm Re}\Pi(m,E_F)$.
The shift is expressed mathematically as follows;
\begin{align}
 & {\rm Re}\Pi(m,E_F) = \theta_{E_F-m} \times \nn \\
 &
 \left\{
 \frac{4\pi E_F}{\hbar \omega_G} - \pi
 \left(
 1+\frac{(2m)^2}{(\hbar\omega_G)^2} \right)
 \ln \left(
 \frac{\hbar\omega_G + 2E_F}{|\hbar\omega_G-2E_F|} \right)
 \right\} +
 \nn \\
 & \theta_{m-E_F} \left\{
\frac{4\pi m}{\hbar \omega_G} - \pi
 \left(
 1+\frac{(2m)^2}{(\hbar\omega_G)^2} \right)
 \ln \left(
 \frac{\hbar\omega_G + 2m}{|\hbar\omega_G-2m|} \right)
 \right\} .
 \label{eq:reP}
\end{align}
When $E_F \le m$, ${\rm Re}\Pi(m,E_F)$ exhibits an anomalous
softening at $m=\hbar \omega_G/2$, which is a logarithmic singularity.
On the other hand, when $m \le E_F$, 
a similar logarithmic singularity is observed for ${\rm Re}\Pi(m,E_F)$
at $E_F=\hbar \omega_G/2$.
The presence of logarithmic singularities on the $m$ and $E_F$ axes
is common. 
However, the factor in front of the logarithm increases with increasing $m$:
$\pi$ ($2\pi$) when $m=0$ ($m=\hbar \omega_G/2$).
The singularity when $m=0$ is referred to as the Kohn anomaly in 
graphene research.~\cite{yan07,pisana07,das08nature,lazzeri06prl}

The behavior of the shift as a function of $m$ differs from that as a
function of $E_F$. 
In particular, ${\rm Re}\Pi(m,E_F)$ converges to zero in the limit of 
$m \to \infty$ (flat), while it increases linearly in proportion to
$E_F$ (hardening).
The plot shows that ${\rm Re}\Pi(m,E_F) \le {\rm Re}\Pi(0,E_F)$ 
meaning that the $G$ band red-shifts when $m$ is increased.
The phonon self-energy is invariant when $|E_F| \le m$ is satisfied, 
since the physical situation when $|E_F| < m$ does not change from
that when $|E_F|=m$.
Finally, the formulas Eqs.~(\ref{eq:-imP}) and (\ref{eq:reP}) can be
used to determine the mass when both $E_F$ and $m$ are dependent
on a sample.

\subsection{The 2D Band}

\begin{figure*}[htbp]
 \begin{center}
  \includegraphics[scale=0.45]{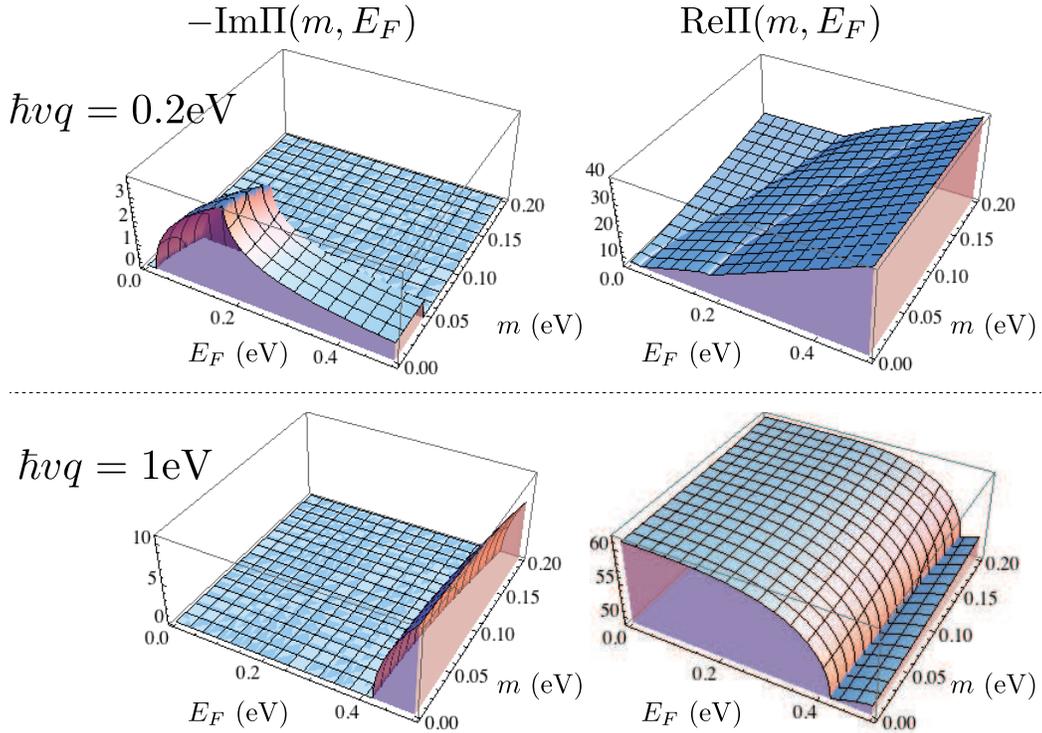}
 \end{center}
 \caption{(color online) The spectrum broadening and shift of the $2D$
 band are given for $\hbar vq=0.2$ and 1 eV, where the phonon energy is
 assumed to be 0.16 eV.
 The broadening for $\hbar vq=0.2$ eV is seen to be sensitive to $m$,
 while the $2D$ Raman spectrum for $\hbar vq=1$ eV is insensitive to mass. 
 All the behavior as a function of $E_F$ for a fixed value of $m$ can be understood by
 shifting the Dirac cones.~\cite{Sasaki2012b}
 }
 \label{fig:2Dpeak}
\end{figure*}

The $2D$ band consists of two intervalley phonons.~\cite{Ferrari2007,Malard2009}
%Meanwhile, the $2D$ band is an overtone of the defect induced $D$ band.~\cite{ferrari06,tuinstra70} 
The wavevector of the intervalley phonon is $2{\bf K}_F+{\bf q}$
where the wavevector $2{\bf K}_F$ is from the the K point to the K$'$ point.
The large intervalley wavevector of the $2D$ band is in contrast to the
vanishing intravalley wavevector of the $G$ band.
Because the $2D$ band has kinematic constraint conditions originating 
from the anisotropy of the electron-phonon interaction, 
the shift $\hbar v|{\bf q}|$ increases with increasing light excitation energy $E_L$
($\hbar v|{\bf q}|\sim 1$ eV when $E_L$ is 1.6 eV).~\cite{Sasaki2012b}
This is known as the dispersive behavior of the $2D$ band.
Since the mass considered in this paper is up to 0.2 eV, which is much
smaller than $\hbar v|{\bf q}|$, 
the idea of shifting a Dirac cone~\cite{Sasaki2012b} 
leads us to understand that the self-energy of the $2D$ band is
generally insensitive to mass.
Only when $\hbar v|{\bf q}|$ (or $E_F$) is of the order of $m$, can the
characteristic signal of the mass appear in the Raman spectrum. 
In Fig.~\ref{fig:2Dpeak}, 
the calculated self-energy of the $2D$
band, $-{\rm Im}\Pi(m,E_F)$ and 
${\rm Re}\Pi(m,E_F)$, is shown 
as a function of $m$ and $E_F$ for $\hbar v q=0.2$ and 1 eV.
These plots are obtained by calculating Eq.~(\ref{eq:selfenergy}) with 
\begin{align}
 |M_{{\bf k+q},{\bf k}}^{s',s}|^2 = 
 \frac{g}{2} \left[
 1 - ss' \left\{ \frac{m^2+(\hbar v)^2 k(k+q\cos\varphi)}{E_{k} E_{|{\bf
 k+q}|}} \right\} \right].
 \label{eq:M2d}
\end{align}
The analytical formula of $-{\rm Im}\Pi(m,E_F)$ is shown below (see
Supplement for ${\rm Re}\Pi(m,E_F)$), which
may be useful when $E_L$ is in the mid infrared region.
\begin{widetext}
\begin{align}
 & -{\rm Im} \Pi(m,E_F) =\pi
 \sqrt{1-\left( \frac{vq}{\omega_D}\right)^2}
 \theta_{\hbar\omega_D -\sqrt{(\hbar vq)^2+(2m)^2}} \nn \\
 & \left[ \pi \theta_{\frac{\hbar \omega-c\hbar vq}{2}-E_F} 
 + \theta_{E_F-\frac{\hbar\omega-c\hbar vq}{2}} \theta_{\frac{\hbar
 \omega+c\hbar vq}{2}-E_F}
 \left\{ \frac{\pi}{2}-\sin^{-1}\left(\frac{2E_F-\hbar\omega_D}{c\hbar vq}\right) \right\}
 \right] + \pi
 \sqrt{\left(\frac{vq}{\omega_D}\right)^2-1} \theta_{vq-\omega_D c} \nn \\
 &
 \Bigg[ \theta_{\frac{c\hbar vq+\hbar \omega_D}{2}-E_F} 
 \theta_{E_F-\frac{c\hbar vq-\hbar\omega_D}{2}}
  g\left( \frac{2E_F+\hbar \omega_D}{c\hbar vq} \right) +
 \theta_{E_F-\frac{\hbar\omega_D+c\hbar vq}{2}}
 \left\{ g\left(\frac{2E_F+\hbar\omega_D}{c\hbar vq}\right)
 -g\left(\frac{2E_F-\hbar \omega_D}{c\hbar vq}\right)\right\}  
 \Bigg],
\end{align}
\end{widetext}
where $c^2=1+\frac{(2m)^2}{(\hbar vq)^2-(\hbar\omega_D)^2}$ and 
$g(x)=\ln(x+\sqrt{x^2-1})$.

The electron-phonon matrix elements of spin-preserved intervalley
scatterings calculated for a topological
insulator are given by replacing $m^2$ with $-m^2$ at 
the numerator on the right side of Eq.~(\ref{eq:M2d}).
The self-energy calculations are possible,
but such extention is beyond the scope of this study.

%%%%%%%%%%%%%%%%%%%%%%%%%%%%
%%%%%%%%%%%%%%%%%%%%%%%%%%%%

\subsection{Domain wall}

It has been reported that the Dirac mass can change its sign spatially 
when the lattice mismatch between graphene and h-BN substrate 
is taken into account.~\cite{Semenoff2008}
Suppose that in Eq.~(\ref{eq:HDirac}) the Dirac mass $m$ ($>0$) 
changes its sign across the $y$-axis at $x=0$ as follows
\begin{align}
 m(x) = m \tanh \left( \frac{x}{\lambda} \right).
\end{align}
It can be shown that a single transport channel with a
topological origin appears along the domain boundary.
To show this we consider first the case where $k_y=0$ in the Hamiltonian of Eq.~(\ref{eq:HDirac}).
The model possesses a single zero-energy state that is normalizable.~\cite{Jackiw1976}
The wavefunction is given by
\begin{align}
 \psi(x) = N \left\{ \cosh\left(\frac{x}{\lambda}\right)
 \right\}^{-\frac{\lambda m}{\hbar v}}
 \begin{pmatrix}
  1 \cr i
 \end{pmatrix},
\end{align}
where $N$ is the normalization constant.
Note that the pseudospin is the positive eigenstate of $\sigma_y$.
The negative eigenstate of $\sigma_y$ is also a zero-energy state.
However, this is not a renormalizable state and therefore must be omitted
from the Hilbert space.

Next, for $k_y\ne 0$, we treat the additional term in the Hamiltonian $\hbar vk_y \sigma_y$ 
as a perturbation and this causes the linear dispersion 
$\hbar vk_y$, since the unperturbed state is the eigenstate
of $\sigma_y$ with an eigenvalue of $+1$.
This mode corresponds to a massless fermion moving in one dimension along the
domain wall at velocity $v$ with the direction of motion
determined by the eigenvalue of $\sigma_y$.
Due to the time-reversal symmetry, two modes appear as a pair (at the K and K$'$ points) and their propagation directions are opposite.
We can expect ballistic massless fermions to be observed, if 
the intervalley backward scattering of the fermions caused by a defect with a short range potential is negligible.
These fermions are invisible unless $E_F$ is in the gap.
Thus, the $E_F$ position in gapped graphene either side of a domain wall is an important issue.
Although we do not expect the $G$ and $2D$ bands to be able to resolve the
domain wall itself, the $G$ band spectrum has sufficient information to
determine the $m$ and $E_F$ values of the gapped graphene, as
we have shown in the preceding sections.

%%%%%%%%%%%%%%%%%%%%%%%%%%%%%
%\begin{figure}[htbp]
% \begin{center}
%  \includegraphics[scale=0.4]{DomainWall.eps}
% \end{center}
% \caption{(color online) Domain wall is formed in gapped graphene when
% the mass changes its sign across the wall (along the $y$-axis).
% The transport of massless fermions at the domain wall is visible when
% $E_F$ is located in the gap.
%}
% \label{fig:DomainWall}
%\end{figure}
%%%%%%%%%%%%%%%%%%%%%%%%%%%%%

\section{Discussion and Conclusion}\label{sec:discussion}

A large band-gap (or heavy mass) $E_g > 0.1$eV ($m>50$meV) may be achieved 
for a monolayer of graphene on h-BN, 
by applying a high pressure to the sample using a diamond anvil cell (DAC).
This speculation is supported by the results described in two papers.
Giovannetti {\it et al}.~\cite{Giovannetti2007} estimated $E_g$ to be larger than 0.1eV theoretically 
when the interlayer distance $d$ between graphene and h-BN is shorter than 3\AA.
They also showed that the ground state of the system is realized when
$d=3.2 \sim 3.5$\AA\ (the equilibrium $d$ values depend on stacking orders).
Meanwhile, a more than 10\% contraction of $d$ was achieved at 10GPa for graphite 
by Hanfland {\it et al}.~\cite{Hanfland1989}
The large contraction along the c-axis is a result of the weak van der
Waals interaction of graphite, which is also expected for graphene on
h-BN heterostructures.
%It is intersting to note that because the contraction along the a-axis of
%graphite is about 1\%, similar a-axis contraction is expected for h-BN,
%but not for monolayer graphene.
%A lattice mismatch between graphene and h-BN substrate causes a strain 
%that can modifies the phonon spectrum.
%The signature of the mass may be appears as the sharpening of the $G$ and $2D$ bands.
%Because the peak positions of the G and 2D depend on the strain effect,
Observing the intensity ratio $I_{2D}/I_{G}$ is useful for confirming the
mass, as well as $E_F$.
When $m=0$, low doping ($E_F < \hbar \omega_G/2$) is confirmed by
observing the maximum $I_{2D}/I_{G}$ 
since the $G$ band broadens while the $2D$ band sharpens.~\cite{das08nature}
Then, $I_{2D}/I_{G}$ increases with increasing $m$ (or pressure)
and $I_{2D}/I_{G}$ reaches its maximum when $m=\hbar \omega_G/2$.
A further increase in $m$ suppresses $I_{2D}/I_{G}$ since 
$I_G$ is suddenly enhanced when $m > \hbar\omega_G/2$.
These signals provide evidence of the mass.

The broadening (or the lifetime) of the $G$ band
has an interesting interpretation as mass-induced orbital motion.
The electron produced by the resonant decay of the $G$ band 
has a classical trajectory with a circular motion whose radius is of the
order of $\hbar v/2m$.
The motion is clockwise or anticlockwise depending on the valley of
the electron. 
This is a result of the time-reversal symmetry of the mass
caused by a staggered sublattice potential.

In summary, we have shown that 
the Raman $G$ band is sensitive to mass because of the pseudospin
polarization of the massive Dirac fermions.
A non-zero Dirac mass can be confirmed by using the self-energy of the
$G$ band through the steep structure of the broadening and flat
structure of the shift.
The $2D$ band is generally insensitive to mass due to for a reason
related to kinematics.
However, there is a possibility that the characteristic
behavior of the mass is visible in the $2D$ band when $E_L$ is
comparable with the energy scale of the mass.
To explore the transport of Weyl fermions at a domain wall, 
the Fermi energy must be close to the charge neutrality.

\section*{Acknowledgments}

We thank H. Sumikura for helpful discussions.

\appendix

\bibliographystyle{apsrev4-1}
%\bibliographystyle{apsrev}
 %\bibliography{/Users/Sasaki/tex/bib/library.bib}
%\bibliography{/Users/sasakikenichi/bib/sasaki,/Users/sasakikenichi/bib/library.bib}

%merlin.mbs 2010-03-15 4.21a (PWD, AO, DPC)

\end{document}